\RequirePackage{fix-cm}
\documentclass{svjour3}                     
\smartqed  
\usepackage{booktabs}
\usepackage{graphicx}
\usepackage{lscape}
\usepackage[table,xcdraw]{xcolor}
\usepackage{listings}
\usepackage{cite}
\usepackage[tight,footnotesize]{subfigure}
\usepackage{xspace}
\usepackage{marvosym}
\usepackage{balance}
\usepackage{url}

\newcommand\etal{\emph{et al.}\xspace}
\newcommand\cparagraph[1]{\vspace{1.5mm}\noindent \textbf{#1}}

\definecolor{mygreen}{rgb}{0,0.6,0}
\journalname{}
\begin{document}

\title{Parallel Programming Models for Heterogeneous Many-Cores : A Survey
}

\titlerunning{Parallel Programming Models for Heterogeneous Many-Cores : A Comprehensive Survey}        

\author{Jianbin Fang         \and
        Chun Huang \Letter  \and
        Tao Tang  \and
        Zheng Wang
}

\authorrunning{J. Fang et. al} 

\institute{J. Fang, C. Huang (\Letter), T. Tang \at
              Institute for Computer Systems, College of Computer, \\ National University of Defense Technology \\
              \email{\{j.fang, chunhuang, taotang84\}@nudt.edu.cn}           
           \and
           Z. Wang \at
              School of Computing, University of Leeds \\
              \email{z.wang5@leeds.ac.uk}
}

\date{Received: date / Accepted: date}

\maketitle

\begin{abstract}

Heterogeneous many-cores are now an integral part of modern computing systems ranging from embedding systems to supercomputers. While
heterogeneous many-core design offers the potential for energy-efficient high-performance, such potential can only be unlocked if the
application programs are suitably parallel and can be made to match the underlying heterogeneous platform. In this article, we provide a
comprehensive survey for parallel programming models for heterogeneous many-core architectures and review the compiling techniques of
improving programmability and portability. We examine various software optimization techniques for minimizing the communicating overhead between
heterogeneous computing devices. We provide a road map for a wide variety of different research areas. We conclude with a
discussion on open issues in the area and potential research directions. This article provides both an accessible introduction to the
fast-moving area of heterogeneous programming and a detailed bibliography of its main achievements.

\keywords{Heterogeneous Computing \and Many-Core Architectures \and Parallel Programming Models}
\end{abstract}

\section{Introduction}







Heterogeneous many-core systems are now commonplace~\cite{DBLP:conf/eurographics/OwensLGHKLP05, DBLP:journals/pieee/OwensHLGSP08}. The
combination of using a host CPU together with specialized processing units (e.g., GPGPUs, XeonPhis, FPGAs, DSPs and NPUs) has been shown in
many cases to achieve orders of magnitude performance improvement. 
As a recent example, Google's Tensor Processing Units (TPUs) are application-specific integrated circuits (ASICs) 
to accelerate machine learning workloads~\cite{DBLP:conf/isscc/Patterson18}. 
Typically, the host CPU of a heterogeneous platform manages the
execution context while the computation is offloaded to the accelerator or coprocessor. Effectively leveraging such platforms not only
enables the achievement of high performance, but increases energy efficiency. These goals are largely achieved using simple, yet
customized hardware cores that use area more efficiently with less power dissipation~\cite{DBLP:journals/ibmrd/ChenRDI07}.

The increasing importance of heterogeneous many-core architectures can be seen from the TOP500 and Green500 list, where a large number of
supercomputers are using both CPUs and accelerators~\cite{top500_url, green500_url}. A closer look at the list of the TOP500
supercomputers shows that seven out of the top ten supercomputers are built upon heterogeneous many-core architectures
(Table~\ref{tab:sc-top10-ppm}). On the other hand, this form of many-core architectures is being taken as building blocks for the
next-generation supercomputers. 
e.g., three US national projects (Aurora~\cite{aurora_url}, Frontier~\cite{frontier_url}, and El Capitan~\cite{capitan_url}) will all implement a
heterogeneous CPU-GPU architecture to deliver its exascale supercomputing systems.

\begin{table*}[]
\caption{The on-node parallel programming models for the top10 supercomputers (as of November 2019).}
\label{tab:sc-top10-ppm}
\resizebox{0.70\textwidth}{!}{%
\begin{tabular}{@{}lclcc@{}}
\toprule
\textbf{Rank} &
  \textbf{Name} &
  \textbf{Compute node architecture} &
  \textbf{Heterogeneous?} &
  \textbf{Programming models} \\ \midrule
\#1 &
  Summit &
  \begin{tabular}[c]{@{}l@{}}IBM POWER9 22C CPU (x2)+\\ NVIDIA Volta GV100 (x6)\end{tabular} &
  YES &
  CUDA/OpenMP \\
\rowcolor[HTML]{C0C0C0}
\#2 &
  Sierra &
  \begin{tabular}[c]{@{}l@{}}IBM POWER9 22C CPU (x2)+\\ NVIDIA Volta GV100 (x4)\end{tabular} &
  YES &
  CUDA/OpenMP \\
\#3 &
  TaihuLight &
  Sunway SW26010 260C &
  YES &
  Athread/OpenACC \\
\rowcolor[HTML]{C0C0C0}
\#4 & Tianhe-2A & \begin{tabular}[c]{@{}l@{}}Intel Xeon E5-2692v2 12C CPU (x2)+\\ Matrix-2000 (x2)\end{tabular}            & YES & OpenCL/OpenMP \\
\#5 &
  Frontera &
  Xeon Platinum 8280 28C CPU &
  NO &
  OpenMP \\
\rowcolor[HTML]{C0C0C0}
\#6 &
  Piz Daint &
  \begin{tabular}[c]{@{}l@{}}Xeon E5-2690v3 12C CPU (x1)+\\ NVIDIA Tesla P100 (x1)\end{tabular} &
  YES &
  CUDA \\
\#7 &
  Trinity &
  \begin{tabular}[c]{@{}l@{}}Intel Xeon E5-2698v3 16C CPU \& \\ Intel Xeon Phi 7250 68C\end{tabular} &
  NO &
  OpenMP \\
\rowcolor[HTML]{C0C0C0}
\#8 & ABCI      & \begin{tabular}[c]{@{}l@{}}Intel Xeon Gold 6148 20C CPU (x2)+\\ NVIDIA Tesla V100 SXM2 (x4)\end{tabular} & YES & CUDA          \\
\#9 &
  SuperMUC-NG &
  Intel Xeon Platinum 8174 24C CPU &
  NO &
  OpenMP \\
\rowcolor[HTML]{C0C0C0}
\#10 &
  Lassen &
  \begin{tabular}[c]{@{}l@{}}IBM POWER9 22C CPU (x2)+\\ NVIDIA Tesla V100 (x4)\end{tabular} &
  YES &
  CUDA/OpenMP \\ \bottomrule
\end{tabular}%
}
\end{table*}

The performance of heterogeneous many-core processors offer a great deal of promise for future computing systems, yet their architecture
and programming model significantly differ from the conventional multi-core processors~\cite{DBLP:conf/sc/BellensPBL06}. This change has
shifted the burden onto programmers and compilers~\cite{DBLP:conf/pldi/KudlurM08}. In particular, programmers have to deal with
heterogeneity, massive processing cores, and a complex memory hierarchy. Thus, programming heterogeneous many-core architectures are
extremely challenging.

How to program parallel machines has been a subject of research for at least four decades~\cite{karp1967organization}. The main contextual
difference between now and the late 80s/early 90s is that heterogeneous parallel processing will be shortly a mainstream activity affecting
standard programmers rather than a high-end elite endeavour performed by expert programmers. This changes the focus from one where raw
performance was paramount to one where programmer productivity is critical. In such an environment, software development tools and
programming models that can reduce programmer effort will be of considerable importance.

In this work, we aim to demystify heterogeneous computing and show heterogeneous parallel programming is a trustworthy and exciting
direction for systems research. We start by reviewing the historical development and the state-of-the-art of parallel programming models for heterogeneous many-cores by
examining solutions targeted at both low-level and high-level programming (Section~\ref{sec:overview}). We then discuss code generation
techniques employed by programming models for improving programmability and/or portability (Section~\ref{sec:bridge}), before turning our
attention to software techniques for optimizing the communication overhead among heterogeneous computing devices (Section~\ref{sec:stream}).
Finally, we outline the potential research directions of heterogeneous parallel programming models (Section~\ref{sec:next}).



\section{Overview of Parallel Programming Models} \label{sec:overview}
\begin{figure}[!b]
	\includegraphics[width=0.52\textwidth]{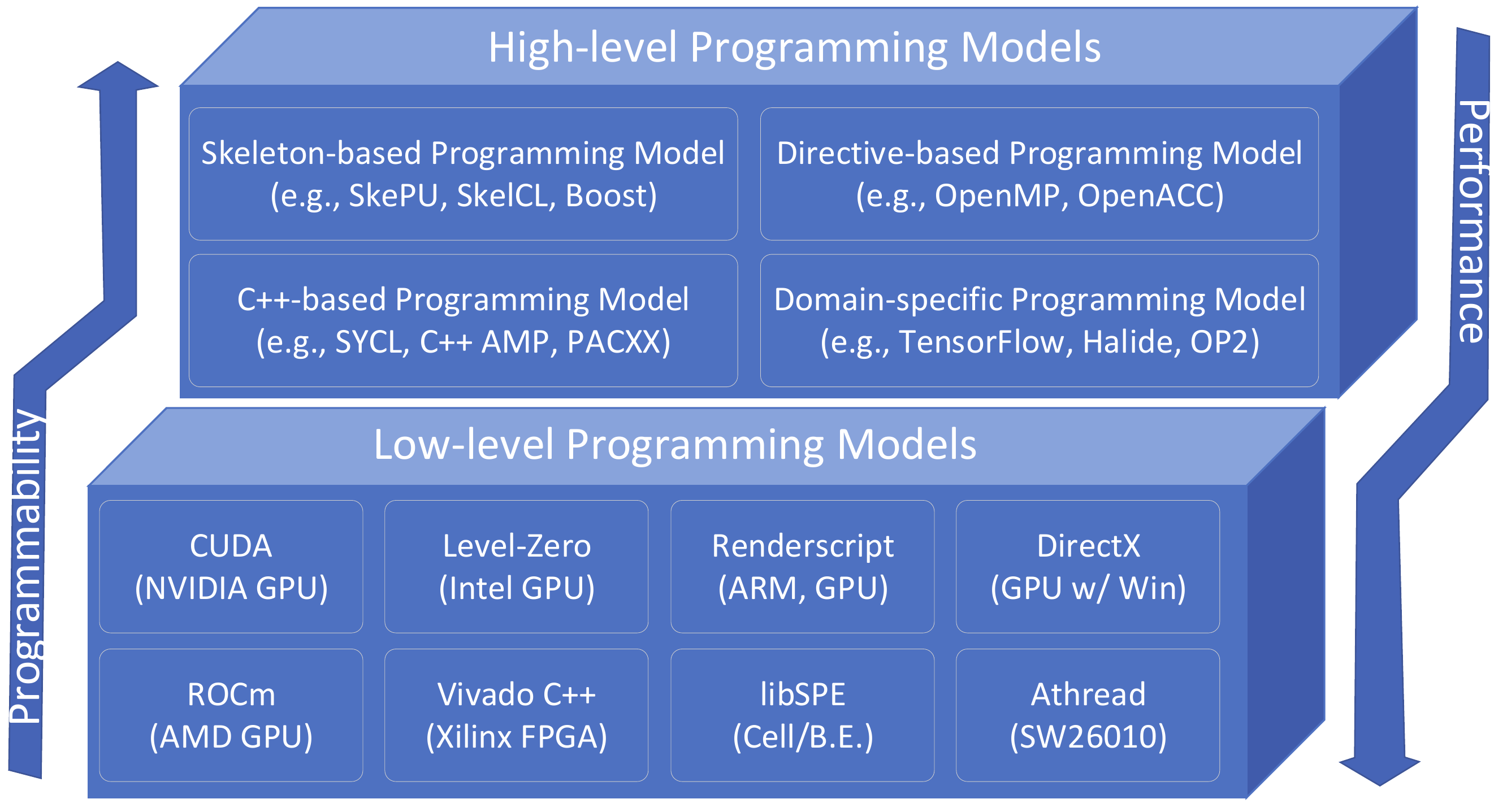}
	\caption{The family of parallel programming models for heterogeneous many-core architectures.}
	\label{fig:ppmodel-family}
\end{figure}
Parallel programming model acts as the bridge between programmers and
parallel architectures.
To use the shared memory parallelism on multi-core CPUs, 
parallel programming models are often implemented on threading mechanisms such as the POSIX threads~\cite{DBLP:conf/usenix/Alfieri94}.
When it comes to heterogeneous many-cores, we have to deal with the heterogeneity between host and accelerators.
And parallel programming models have to introduce relevant abstractions of controlling them both,
which is the focus of this survey work. 

Figure~\ref{fig:ppmodel-family} summarizes
the family of parallel programming models for heterogeneous many-core architectures.
Based on the performance-programmability tradeoff,
we categorize them into
\textit{low-level programming models} (Section~\ref{sec:lowlevel-ppm}) and \textit{high-level programming models} (Section~\ref{sec:highlevel-ppm}).
The expected application \textit{performance} increases from high-level programming models to low-level programming models, 
whereas the \textit{programmability} decreases.

The low-level programming models are closest to the many-core architectures, and
expose the most hardware details to programmers through data structures and/or APIs.
These models are typically bound to specific hardware architectures, and are also
known as \textit{native programming models}.
In contrast, the high-level programming models raise the languages' abstraction level,
and hide more architecture details than the low-level models. 
Thus, the high-level models often enable better programmability.


\subsection{Low-level Parallel Programming Models} \label{sec:lowlevel-ppm}
\vspace{-2mm}
\subsubsection{Prelude of GPGPU Programming}
\vspace{-2mm}
\textbf{The GPU Shading Lanuages} At early 2000s, commodity graphics hardware was rapidly evolving from a fixed function pipeline into
a programmable vertex and fragment processor.
Originally, these programmable GPUs could only be programmed using assembly language.
Later, Microsoft and NVIDIA introduced their C-like programming languages, HLSL and Cg respectively, that compile to
GPU assembly language~\cite{hlsl_url, DBLP:journals/tog/MarkGAK03}.
The shading languages make it easier to develop programs incrementally and interactively.
This is achieved by using a high-level shading language, e.g., Cg, based on both the syntax and the philosophy of C.


Although these shading languages can hide details of the graphics pipeline (e.g., the number of stages or
the number of passes), they are specialized for real-time shading and remain very graphics-centric~\cite{DBLP:journals/tog/BuckFHSFHH04}.
In particular, these high-level shading languages
do not provide a clean abstraction for general-purpose computing on graphics hardware.
Programs operate on vertices and fragments separated by a rasterization stage; memory is divided up into
textures and framebuffers; the interface between the graphics hardware and host is through a complex graphics API.
This often prevents applying the graphics hardware onto new applications.

\cparagraph{Brook for GPGPUs}
The graphics processors feature instruction sets
general enough to perform computation beyond the rendering domain.
Applications such as linear algebra operators~\cite{DBLP:journals/tog/KrugerW03},
numerical simulation~\cite{DBLP:conf/egh/HarrisBSL03},
and machine learning algorithms~\cite{DBLP:conf/icdar/SteinkrauSB05} have been ported to GPUs and achieved a remarkable speedup over traditional CPUs.
These research works demonstrate the potential of graphics hardware for more general-purpose computing tasks, i.e., GPGPUs.



The first work to explore this idea is the \texttt{Brook} programming system~\cite{DBLP:journals/tog/BuckFHSFHH04}.
By introducing the concepts of \textit{streams}, \textit{kernels} and \textit{reduction operators},
Brook abstracts the GPU as a streaming processor. This abstraction is achieved by
virtualizing various GPU hardware features with a compiler and runtime system.

The Brook language is an extension to the standard ANSI C and is designed to
incorporate the idea of data parallel computing and arithmetic intensity.
A Brook program consists of legal C code plus syntactic extensions to denote streams and kernels.
The Brook programming system
consists of \texttt{BRCC} and \texttt{BRT}.
\texttt{BRCC} is a source-to-source compiler which translates Brook codes (\texttt{.br}) into C++ codes (\texttt{.cpp}).
\texttt{BRT} is a runtime software which implements the backend of the Brook primitives for target hardware.
\textit{We regard Brook as
an origin work for programming GPGPUs, and
other parallel programming models for heterogeneous many-cores inherit many features from it. }

\subsubsection{Vendor-Specific Programming Models} \label{subsec:vendor-ppm}
Vendor-specific programming models are bound to vendors and their manufactured hardware.
The typical heterogeneous many-core architectures include Cell/B.E., NVIDIA GPU,
AMD GPU, Intel XeonPhi, FPGA, DSP, and so on~\cite{DBLP:journals/sp/BrodtkorbDHHS10}.
Hardware vendors introduces their unique programming interfaces, which are restricted to their own products.
This section examines each many-core architecture and its native programming models.

\cparagraph{libSPE for IBM Cell Broadband Engine}
The Cell Broadband Engine Architecture (CEBA) and its first implementation Cell/B.E.
is a pioneering work of heterogeneous computing~\cite{DBLP:journals/ibmrd/KahleDHJMS05, DBLP:journals/micro/GschwindHFHWY06}.
Cell/B.E.~\cite{DBLP:conf/cicc/PhamBBHJKKKLMPR05} was designed by
a collaboration effort between Sony, Toshiba, and IBM (STI),
and takes a radical departure from conventional multi-core architectures.
Instead of integrating identical commodity hardware cores, it uses a conventional high performance PowerPC core (PPE) which
controls eight simple SIMD cores (i.e., Synergistic Processing Elements, SPEs).
Each SPE contains a synergistic processing unit (SPU), a local store, and a memory flow controller (MFC).
Prior works have demonstrated
that a wide variety of algorithms on the Cell/B.E. processor can achieve performance
that is equal to or significantly better than a general-purpose
processor~\cite{DBLP:conf/cf/WilliamsSOKHY06, DBLP:journals/ibmrd/ChenRDI07, DBLP:conf/hipc/BaderA07, DBLP:journals/tpds/ScarpazzaVP08}.
Its architecture variant, i.e., \texttt{PowerXCell 8i}, has been used to build the first peta-scale supercomputer,
\texttt{Roadrunner}~\cite{roadrunner_book, DBLP:conf/sc/BarkerDHKLPS08, DBLP:conf/ppopp/KistlerGBB09}.

Programming the CEBA processor is challenging~\cite{DBLP:conf/cf/CrawfordHKW08}.
IBM developed IBM SDK for Multicore Acceleration with a suite of software tools and libraries~\cite{cell_book}.
The SDK provides various levels of abstractions, ranging from the low-level management library to the high-level programming models,
to tap the Cell's potential.
Managing the code running on the SPEs of a CEBA-based system can be
done via the \texttt{libspe} library (SPE runtime management library) that is part of the
SDK package~\cite{cell_book}.
This library provides a standardized low-level programming interface that manages the SPE threads,
and enables communication and data transfer between PPE threads and SPEs.
Besides, the SDK contains high-level programming frameworks to assist the development of parallel applications on this architecture.

\cparagraph{CUDA for NVIDIA GPUs}
NVIDIA implemented the unified shader model, where all shader units of graphics hardware
are capable of handling any type of shading tasks,
in the Tesla and its subsequent designs~\cite{csc_url}.
Since G80, NVIDIA's GPU architecture has evolved from Tesla~\cite{DBLP:journals/micro/LindholmNOM08, DBLP:conf/ispass/WongPSM10},
Fermi~\cite{nvidia-fermi-architecture-whitepaper}, Kepler~\cite{nvidia-kepler-architecture-whitepaper},
Maxwell~\cite{nvidia-maxwell-architecture-whitepaper},
Pascal~\cite{nvidia-pascal-architecture-whitepaper}, Volta~\cite{nvidia-volta-architecture-whitepaper},
to Turing~\cite{nvidia-turing-architecture-whitepaper}.
Each generation of NVIDIA's microarchitecture introduces new features
based on its previous one,
e.g., the Volta architecture features tensor cores that have superior
deep learning performance over regular CUDA cores.

NVIDIA introduces CUDA to program its computing architecture for general-purpose computation~\cite{cuda_url}.
The CUDA programming model works with programming languages such as C, C++, and Fortran.
A CUDA program calls parallel kernels, with each executing in parallel across a set of
threads. The programmer organizes these threads in thread blocks and grids of
thread blocks. The GPU instantiates a kernel program on a grid of parallel thread blocks.
Each thread within a thread block executes an instance of the kernel, and has a thread ID
within its thread block, program counter, registers, and per-thread private memory.
This accessibility makes it easier for us to use GPU resources.

With CUDA, NVIDIA GPUs have been used to speed up
both regular applications~\cite{DBLP:journals/pc/TomovDB10, DBLP:conf/sc/GovindarajuLDSM08}
and irregular ones~\cite{DBLP:conf/ppopp/HongKOO11, DBLP:conf/ppopp/MerrillGG12},
with an impressive performance increase over multi-core CPUs.
Nevertheless, Lee \etal argue that the performance gap can be narrowed by applying optimizations
for both CPUs and GPUs~\cite{DBLP:conf/isca/LeeKCDKNSSCHSD10}.
Table~\ref{tab:sc-top10-ppm} shows that five of the top ten supercomputers use NVIDIA GPUs as the accelerators.
The GPU-enabled architectures will continue
to play a key role in building future high-performance computing systems.


\cparagraph{CAL/ROCm for AMD GPUs}
AMD/ATI was the first to implement the unified shader model in its TeraScale design,
leveraging flexible shader processors which can be scheduled to process a variety of shader types~\cite{csc_url}.
The TeraScale is based upon a very long instruction word (VLIW) architecture, in which the core executes operations in parallel.
The Graphics Core Next (GCN) architecture moved to a RISC SIMD microarchitecture,
and introduced asynchronous computing~\cite{amd-gcn-vega-architecture-whitepaper}.
This design makes the compiler simpler and leads to a better utilization of hardware resources.
The RDNA (Radeon DNA) architecture is optimized for efficiency and programmability, while offering
backwards compatibility with the GCN architecture~\cite{amd-rdna-architecture-whitepaper}.
As the counterpart to the gaming-focused RDNA, CDNA is AMD's compute-focused architecture for HPC and ML workloads.

Close-To-the-Metal (CTM) is a low-level programming framework for AMD's GPUs.
This framework enables programmers to access AMD GPUs with a high-level abstraction, Brook+~\cite{amd-brookplus-programming},
which is an extension to the Brook GPU specification on AMD's compute abstraction layer (CAL)~\cite{amd-cal-programming}.
Then AMD renamed the framework as AMD APP SDK (Accelerated Parallel Programming) built upon AMD's CAL, with an OpenCL programming interface.
In November 2015, AMD released its ``Boltzmann Initiative" and the ROCm open computing platform~\cite{rocm_url}.
ROCm has a modular design which lets any hardware-vendor drivers support the ROCm stack~\cite{rocm_url}.
It also integrates multiple programming languages, e.g., OpenCL and HIP, and
provides tools for porting CUDA codes into a vendor-neutral format~\cite{hip_url}.
At the low level, ROCm is backed by a HSA-compliant language-independent runtime, which resides on the kernel driver~\cite{rocmr_url}.

On AMD GPUs,
a kernel is a single sequence of instructions that operates on a
large number of data parallel work-items. The work-items are organized into architecturally
visible work-groups that can communicate through an explicit local data share (LDS). The
shader compiler further divides work-groups into microarchitectural wavefronts that are
scheduled and executed in parallel on a given hardware implementation.
Both AMD and NVIDIA use the same idea to hide the data-loading latency and achieve high throughput,
i.e., grouping multiple threads.
AMD calls such a group a \textit{wavefront}, while NVIDIA calls it a \textit{warp}.



\cparagraph{MPSS/COI for Intel XeonPhis}
Intel XeonPhi is a series of x86 manycore processors,
which inherit many design elements from the Larrabee project~\cite{DBLP:journals/micro/SeilerCSFDJLCEGJASH09}.
It uses around 60 cores and 30 MB of
on-chip caches, and features a novel 512-bit vector processing unit within a core~\cite{DBLP:conf/wosp/FangSZXCV14}.
This architecture has been used to build the Tianhe-2 supercomputer,
which was ranked the world's fastest supercomputer in June 2013~\cite{th2_url}.




The main difference between an XeonPhi and a GPU is that XeonPhi
can, with minor modifications,
run software that was originally targeted to a standard x86 CPU.
Its architecture allows the use of standard programming languages and APIs such as OpenMP.
To access the PCIe-based add-on cards,
Intel has developed the Manycore Platform Software Stack (MPSS) and the Coprocessor Offload Infrastructure (COI)~\cite{mpss_url}.

Intel COI is a software library designed to ease the development of software and applications
that run on Intel XeonPhi powered device.
The COI model exposes a pipelined programming model, which
allows workloads to be run and data to be moved asynchronously.
Developers can configure one or more pipelines to interact between sources and sinks.
COI is a C-language API that interacts with workloads through standard APIs.
It can be used with any other programming models, e.g., POSIX threads.


\begin{figure}[!t]
  \centering
  \includegraphics[width=0.45\textwidth]{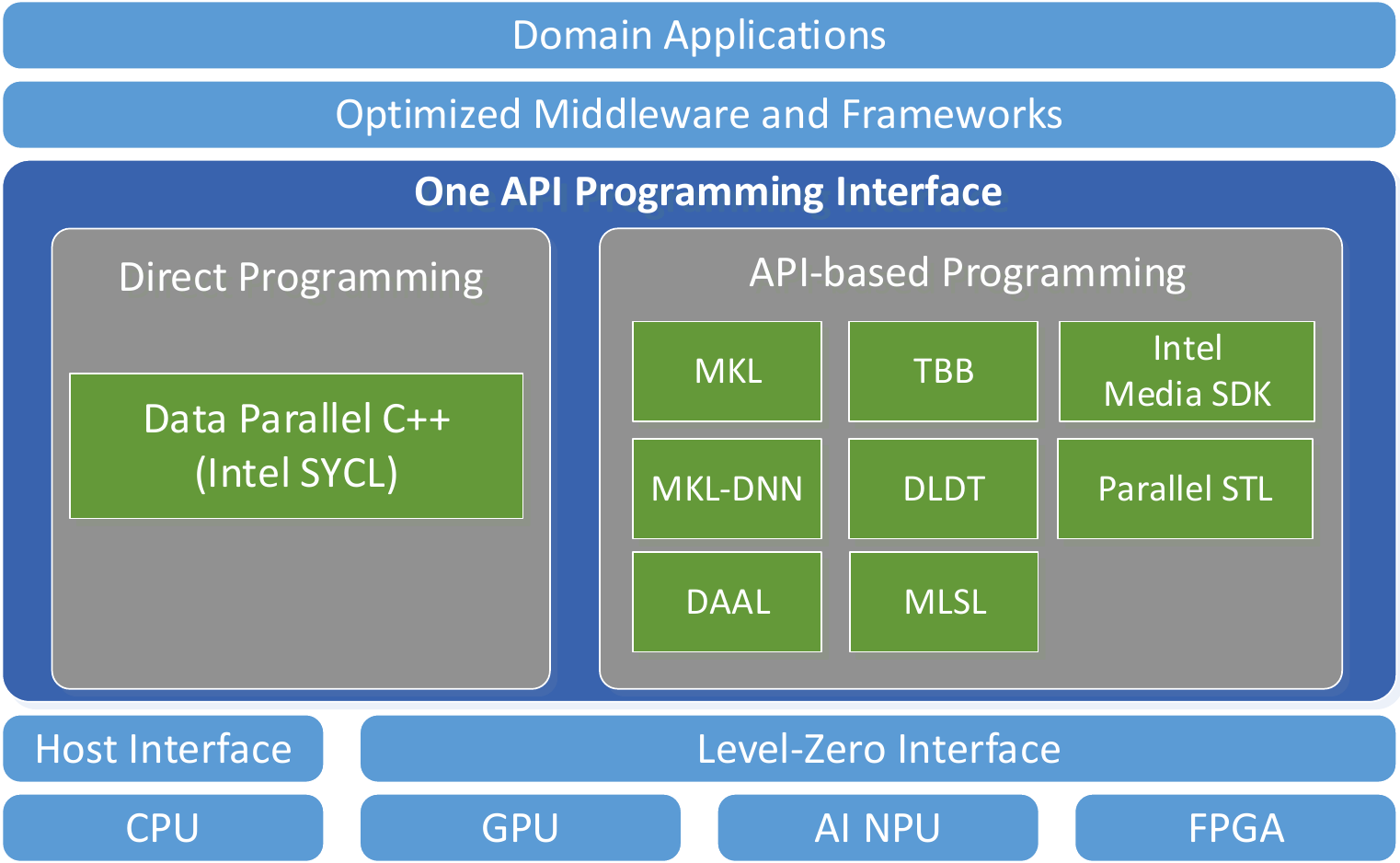}
  \caption{Intel's OneAPI software stack. Reproduced from~\cite{oneapi_url}.}
  \label{fig:one-api-sw-stack}
\end{figure}

\cparagraph{Level-Zero for Intel XPUs} Intel GPUs have been getting more powerful
and are strong contenders in the graphics and GPGPU space~\cite{DBLP:conf/cgo/ChandrasekharCC19}.
Apart from the integrated graphics, Intel has revealed its discrete GPU architecture (Xe)
targeted for datacenter and HPC applications, e.g., being as the primary compute engine for the Aurora supercomputer~\cite{aurora_url}.
Figure~\ref{fig:one-api-sw-stack} shows Intel's \texttt{OneAPI} software stack
to unify programming across its compute product portfolio (CPUs, GPUs, NPUs and FPGAs)
with a single set of APIs~\cite{oneapi_url}.
At the low level, the \texttt{Level-Zero} API is to provide direct-to-metal interfaces to offload accelerator devices.
This interface is intended for providing explicit controls needed by higher-level runtime APIs and libraries, e.g., \texttt{OneAPI}.
Its design is initially influenced by other low-level APIs.

\cparagraph{DirectCompute for GPUs}
Microsoft's DirectCompute is an application programming interface that supports running compute kernels
on GPGPUs
on Windows.
DirectCompute is part of the Microsoft DirectX collection of APIs.

DirectCompute exposes GPU's compute functionality as a new type of shader -
the compute shader, which is very similar to the existing vertex, pixel and geometry
shaders, but with much more general purpose processing capabilities~\cite{microsoft-directcompute}.
The compute shader is not attached specifically to any stage of the graphics pipeline, but
interacts with the other stages via graphics resources such as render targets, buffers and
textures. Unlike a vertex shader, or a pixel shader, the compute shader has no fixed
mapping between the data it is processing and threads that are doing the processing.
The compute shader also allows unordered memory access, in particular the ability to
perform writes to any location in a buffer.
The DirectCompute architecture shares a range of computational interfaces with NVIDIA's CUDA.

\cparagraph{Renderscript for GPUs}
Many mobile phones have followed the same
trend of desktop architectures, integrating different types of
processing units onto the same chip.
These new mobile phones include multi-core
CPUs (e.g., Qualcomm Snapdragon)
as well as GPUs
(e.g., ARM Mali or Qualcomm Adreno).


Google released Renderscript as an official GPU computing
framework for Android in 2011~\cite{renderscript_url}. Renderscript provides
high performance and portability across mobile hardware architectures.
It provides three primary tools:
a simple 3D rendering API, a compute API similar to CUDA, and a C99-derived language.

\subsubsection{A Unified Programming Model}


The use of heterogeneous many-core architectures in high-performance computing has attracted increasingly more interests,
particularly due to the growth of graphics processing units.
Much of this growth has been driven by NVIDIA's CUDA ecosystem for developing GPGPU applications on NVIDIA GPUs.
However, with the increasing diversity of GPUs, including those from AMD, ARM, Intel and Qualcomm,
OpenCL (Open Computing Language) has emerged as an open and vendor-agnostic standard for programming GPUs as well as other accelerating devices
such as APUs and FPGAs.

OpenCL is an open programming standard that is maintained by the Khronos group.
Its API consists of a C library supporting device programming in the C99 or C++ language.
An OpenCL application is composed of two parts: one or more kernels
and an OpenCL host program.
The kernel
specifies functions to be executed in a parallel fashion on
the processing cores. The host sets up the execution
environment, reads in data, and instantiates and enqueues the
kernels to run.

\cparagraph{Code Portability vs. Performance Portability}\\
OpenCL stands out in its portability by defining an abstract execution model and a platform model.
Porting OpenCL to a new many-core device is a matter of providing an implementation of
the runtime library that conforms to the standard, achieving the goal of code portability~\cite{DBLP:phd/basesearch/Fang14}.
OpenCL applications written for one vendor's platform should run correctly on other vendors' platforms, if
they are not using any vendor-proprietary or platform-specific extensions.
The code portability of OpenCL is ensured by Khronos' certification program, which requires
OpenCL vendors to pass rigorous conformance tests on their platform before they claim it is
OpenCL ``conformant''~\cite{qualcomm_ocl_tr, opencl_cts_url}.

Different from code portability, OpenCL cannot guarantee the goal of performance portability.
This is because
the hardware implementation of OpenCL is vendor dependent.
Different hardware vendors have their unique device architectures.
As a result, an OpenCL application written and optimized for one vendor's
platform is unlikely to have the same performance as on other vendors' platforms~\cite{bodin:hal-00920910}.
To achieve portable performance, researchers have investigated various techniques,
which are discussed in Section~\ref{sec:bridge}.



\cparagraph{Closed and Open-source Implementation}
Table~\ref{tab:ocl-implement} shows that there exist a variety of OpenCL implementations.
AMD is working on the ``Boltzmann Initiative'' and the ROCm open compute platform,
which contains its OpenCL implementation~\cite{amd_opencl_url}.
Furthermore, the Gallium Compute Project
maintains an implementation of OpenCL mainly for AMD
Radeon GCN (formerly known as CLOVER~\cite{gallium_opencl_url}), and it
builds on the work of the Mesa project to support multiple
platforms.
Recently, Intel has turned to implementing OpenCL for its CPUs, GPUs and FPGAs,
and made its partial implementation open to the public~\cite{intel_opencl_url}.
BEIGNET was an open-source implementation of OpenCL released by
Intel in 2013 for its GPUs (Ivy Bridge and newer), but is now deprecated~\cite{beignet_opencl_url}.
IBM once released its OpenCL implementation for
programming CEBA~\cite{DBLP:conf/ipps/BreitbartF10}.

\begin{table*}[!t]
\centering
\caption{The OpenCL Implementations: Open- and closed-source}
\label{tab:ocl-implement}
\resizebox{0.99\textwidth}{!}{%
\begin{tabular}{@{}llllllc@{}}
\toprule
                & Developer   & SDK            & Hardware                     & Operating System & Version  & Open-Source \\ \midrule
AMD OpenCL      & AMD         & ROCm           & AMD CPU/GPU/APU              & Linux/Windows    & 2.0      & Y           \\
\rowcolor[HTML]{C0C0C0}
NVIDIA OpenCL   & NVIDIA      & CUDA           & NVIDIA GPU                   & Linux/Windows    & 1.2      & N           \\
Intel OpenCL    & Intel       & Intel SDK      & Intel CPU/GPU/MIC/FPGA       & Linux/Windows    & 2.1      & Y           \\
\rowcolor[HTML]{C0C0C0}
IBM OpenCL      & IBM         & IBM SDK        & IBM CPU/CEBA                 & Linux            & 1.1      & N           \\
ARM OpenCL      & ARM         & ARM            & ARM CPU/Mali GPU             & Linux            & 1.2, 2.0 & N           \\
\rowcolor[HTML]{C0C0C0}
Qualcomm OpenCL & Qualcomm    & Adreno GPU SDK & Qualcomm Adreno GPU          & Andriod          & 2.0      & N           \\
TI OpenCL       & TI          & Processor SDK  & TI C66x DSP                  & Linux            & 1.1      & Y           \\
\rowcolor[HTML]{C0C0C0}
ZiiLABS OpenCL  & ZiiLABS     & ZIILABS SDK    & ZMS StemCell processors      & Andriod          & N/A      & N           \\
POCL            & Tampere U.  & POCL           & CPU/ASIP/NVIDIA GPU/HSA GPUs & Linux/Windows    & 1.2, 2.0 & Y           \\
\rowcolor[HTML]{C0C0C0}
Clover OpenCL   & Denis Steck & Mesa           & AMD GPU                      & Linux            & 1.1      & Y           \\
FreeOCL         & zuzuf       & FreeOCL        & CPU                          & Linux/Windows    & 1.2      & Y           \\
\rowcolor[HTML]{C0C0C0}
MOCL            & NUDT        & MOCL           & Matrix-2000                  & Linux            & 1.2      & Y           \\
SnuCL           & SNU         & SNUCL          & CPU/GPU/Cluster              & Linux            & 1.2      & Y           \\
\bottomrule
\end{tabular}%
}
\end{table*}

In recent years, the mobile system-on-chips (SOCs) have advanced significantly in computing
power. GPUs in the mobile SOCs are very
powerful.
To leverage the computing power, mobile vendors enable OpenCL onto their hardware.
ZiiLABS enabled OpenCL on the ZiiLABS platforms and released the ZiiLABS OpenCL SDK~\cite{ziilabs_url}.
This implementation aims to unlock the full potential of the StemCell (SIMD) array architecture to
deliver new levels of performance.
ARM has also implemented OpenCL for Mali GPUs~\cite{DBLP:conf/ipps/GrassoRRGR14}.
Qualcomm provides the Qualcomm Adreno SDK to take full advantage of
the graphics and computation power provided by the Adreno GPU~\cite{qualcomm_ocl_tr}.
To facilitate seamless migration of applications between TI SoCs,
TI has customized OpenCL implementation for its SOCs (ARM CPUs+TI DSP)~\cite{texasinstruments_ocl_url}.



There are several open-source implementations developed
and maintained by the academia.
\texttt{POCL} is an implementation built on Clang and
LLVM.
It supports
CPUs, TTA, NVIDIA GPUs, and the HSA-based architectures~\cite{DBLP:journals/ijpp/JaaskelainenLSR15}.
Based on POCL, the researchers from National University of Defense Technology have built an OpenCL implementation (\texttt{MOCL})
for the Matrix-2000 many-core architecture~\cite{DBLP:conf/cf/ZhangTFHYW18} and the FT-1500A multi-core CPU~\cite{DBLP:conf/ispa/FangZTHY17}.
With the help of the generic C++ compilers, \texttt{FreeOCL} can supports a large range of multi-core CPUs~\cite{free_opencl_url}.
\texttt{SnuCL} is an open-source OpenCL framework developed at Seoul National University.
This framework stands out that it extends the original OpenCL semantics to the heterogeneous cluster environment~\cite{DBLP:conf/ics/KimSLNJL12}.

\cparagraph{One OpenCL to Rule them All?}
It has been around ten years since the birth of the OpenCL standard in 2009~\cite{ocl_url, DBLP:conf/icpads/MartinezGF11}.
We had expected that, OpenCL became the unified de facto standard programming model for heterogeneous many-core processors,
like OpenMP for multi-core CPUs~\cite{omp_url} and MPI for large-scale clusters~\cite{mpi_url}.
However, this is not the case.
The fact is that, CUDA has been the de facto programming environment on GPGPUs for years.
Table~\ref{tab:sc-top10-ppm} shows that five of the top ten supercomputers
use GPGPU architectures and take CUDA as their programming methodology.

We believe that there are several factors behind OpenCL's tepid popularity.
The first factor is due to the diversity of many-core architectures in terms of processing cores and memory hierarchies.
For example, using scratch-pad memory has been very popular in DSPs, game consoles (IBM Cell/B.E.), and graphic processor,
while caches are typically used in Intel XeonPhi.
To be compatible with the diverse many-cores, the next version of the OpenCL standard will
let more OpenCL features become optional for enhanced deployment flexibility.
The optionality includes both API and language features e.g., floating point precisions.
By doing so, we can enable vendors to ship functionality precisely targeting their customers and markets~\cite{opencl-architecture-update-whitepaper}.
The second factor is due to the commercial interests.
The vendors would prefer using a specialized programming model for their hardware and building a complete software ecosystem.
From the perspective of programmers, we often choose to use the CUDA programming interface for NVIDIA GPU.
This is because NVIDIA optimize its CUDA software stack, and
thus applications in CUDA can often enable us to yield a better performance~\cite{DBLP:conf/icpp/FangVS11}.



To summarize, \textit{OpenCL is eligible, yet not practical to be a unified programming model for heterogeneous many-core architectures}.
For future, we argue that, neither OpenCL nor CUDA dominates the programming range for heterogeneous many-core architectures,
but it is likely that they coexist.
Our investigation work shows that the low-level programming models share a common working paradigm,
and vendors would choose to support their unique one.
Thereafter, they should develop a complete ecosystem (e.g., domain libraries) around this programming model.

\subsection{High-level Parallel Programming Models} \label{sec:highlevel-ppm}
To lower programming barrier,
various strategies have been proposed to improve the abstraction level,
i.e., high-level programming models.
This is achieved by redefining programmers' role
 and
letting compilers and runtime systems take more work.
Ultimately, the high-level parallel programming models aim to
free programmers from mastering the low-level APIs
or performing architecture-specific optimizations.
There are five types of high-level programming models:
(1) the C++-based programming models,
(2) the skeleton-based programming models,
(3) the STL-based programming models,
(4) the directive-based programming models, and
(5) the domain-specific programming models.

\subsubsection{C++-based Programming Models}
SYCL is a cross-platform abstraction layer that builds on OpenCL's concepts, portability and efficiency for programming heterogeneous platforms
in a single-source style with standard C++.
This programming model enables the host and kernel code
for an application to be contained in the same source file,
and achieves the simplicity of a cross-platform asynchronous task graph~\cite{sycl-spec}.
Intel has been developing oneAPI that includes DPC++ (an implementation of SYCL with extensions) for its CPUs, GPUs and FPGAs~\cite{oneapi_url}.
C++ AMP (C++ Accelerated Massive Parallelism) is a heterogeneous programming model based on C++.
Built upon DirectX11, this model uses \textit{parallel\_for\_each} to instantiate device code, and provides users with a C++ programming interface.
Thus, the programmer can write GPU code in a more concise and controlled way~\cite{amp_book}.
Inspired by C++ AMP and C++14, HC (Heterogeneous Computing) C++ API is a GPU-oriented C++ compiler developed by AMD~\cite{hcc_url}.
HC removes the ``restrict'' keyword,
supports additional data types in kernels, and
provides fine-grained control over synchronization and data movement.

\texttt{PACXX} is a unified programming model for programming heterogeneous many-core systems,
to mitigate the issue of long, poorly structured and error-prone codes in low-level programming models~\cite{DBLP:conf/sc/HaidlG14}.
In PACXX, both host and device programs are written in the
C++14 standard,
with all modern C++ features including type inference (auto),
variadic templates, generic lambda expressions, as well as
STL containers and algorithms.
PACXX consists of a custom compiler (based on the Clang front-end and
LLVM IR) and a runtime system,
which facilitate automatic memory management and data synchronization~\cite{DBLP:journals/ijpp/HaidlG18, DBLP:conf/sc/HaidlMKSHG17}.

Other C++-based parallel programming models include boost.compute~\cite{DBLP:conf/iwocl/Szuppe16},
HPL~\cite{DBLP:journals/jpdc/VinasBF13, DBLP:journals/concurrency/VinasFAD18}, VexCL~\cite{denis_demidov_2017_571466},
hlslib for FPGAs~\cite{DBLP:journals/corr/abs-1910-04436}, alpaka~\cite{DBLP:conf/ipps/ZenkerWWHJKNB16}, and so on.
By respecting the philosophy of modern C++,
these high-level programming models integrate
the host code and the device code into a single C++ file,
and use a unified data structure to manage the buffer to avoid manual management of data synchronization.
By doing so, programmers can transparently use the parallel computing resources of many-core architectures,
without having to master their details.
An important feature of the C++-based parallel programming model is that,
it can form a natural match with other C++ programs, which facilitates the development and composition of
large-scale scientific computing applications, significantly improving programmability and productivity.

\subsubsection{Skeleton-based Programming Models\label{sec:skeleton}}

Skeleton programming is a programming paradigm based on algorithmic skeleton~\cite{skeleton_book}.
A skeleton is a predefined ``high-order function'', such as \textit{map}, \textit{reduce}, \textit{scan}, \textit{farm}, and \textit{pipeline},
which implements a common pattern of computation and/or data dependence.
Skeleton programming provides a high level of abstraction and portability,
with a quasi-sequential programming interface,
as their implementations encapsulate all the underlying details and architecture features,
including parallelization, synchronization, communication, buffer management, accelerator usage and many others.


SkePU is a skeleton programming framework based on the C++ template for multi-core CPU and multi-GPU systems.
This framework contains six data-parallel and one task-parallel skeletons, and two generic container types.
The backend support of SkePU includes OpenMP, OpenCL and CUDA,
where Clang is used to facilitate the source-to-source
code transformation~\cite{DBLP:conf/appt/DastgeerLK13}.
The interface of SkePU2 is improved based on C++11 and variadic templates,
and user-defined functions are expressed with the lambda expression~\cite{DBLP:journals/ijpp/ErnstssonLK18}.

SkelCL is another skeleton library for heterogeneous platforms~\cite{DBLP:conf/ipps/SteuwerKG11}.
It provides programmers with vector data type, high-level data distribution mechanism to
enable the automatic buffer management, implicit data transfers between host and accelerator.
This aims to significantly simplify the programming of heterogeneous many-core systems.
The SkelCL backend translates each skeleton into an OpenCL kernel, and
enables the SkelCL codes to run on both multi-core CPUs and heterogeneous GPUs.
This framework also supports automatic mapping of tasks onto multiple devices~\cite{DBLP:journals/tjs/SteuwerG14}.

Other skeleton-based programming models include Muesli~\cite{muesli_techreport}, Marrow~\cite{DBLP:conf/europar/MarquesPAM13},
ParallelME~\cite{DBLP:conf/europar/AndradeCUCAFRFG16, DBLP:conf/sbac-pad/MoreiraACGFRSA17}, etc.
Compared with the C++-based programming models, the skeleton programming model is lack of generality,
i.e.,  some computing or communication patterns are difficult to be represented
by the builtin skeletons.
Thus we have to extend the existing skeletons
when necessary.
Programmers are also responsible for synthesizing their target applications with builtin skeletons.


\subsubsection{STL-based Programming Models}
TBB (Threading Building Blocks) is a C++ template library developed by Intel for parallel programming of its multi-core CPUs.
It implements a set of algorithm components (e.g., \textit{parallel\_for}, \textit{parallel\_reduce}) and
a set of containers (e.g., \textit{concurrent\_queue}, \textit{concurrent\_vector}).
The TBB runtime system is responsible for managing and scheduling threads to execute tasks, and
balancing the computing loads between multi-cores by task stealing.
By doing so, TBB aims to unbind programmers and the underlying hardware~\cite{DBLP:journals/software/KimV11}.
HPX (High Performance ParallelX) is a generic C++ runtime system for parallel and distributed systems~\cite{hpx_su, DBLP:conf/sc/HellerKSF13}.
By providing a unified programming interface,
HPX can transparently use the underlying hardware resources with an asynchronous multi-tasking mechanism.
HPX aims to be easy-to-use, and achieves high scalability and portability.
HPX enables the support of heterogeneous computing,
by introducing the concepts of \textit{target}, \textit{allocator} and \textit{executor} within the \texttt{hpx.compute} subproject.
The backend of the computing platform includes CUDA, HCC and SYCL~\cite{DBLP:conf/iwocl/CopikK17, DBLP:conf/supercomputer/HellerKDFS16}.

Thrust is a C++ standard parallel template library for NVIDIA GPUs.
It implements four basic core algorithms for \textit{each}, \textit{reduce}, \textit{scan} and \textit{sort}~\cite{BELL2012359}.
By doing so, users need not know how to map the calculation to the computing resources
 (e.g., thread block size, buffer management, algorithm variant selection, etc.),
but only pay attention to the calculation itself.
This approach can greatly improve productivity.
Thrust's backend is based on CUDA, TBB, and OpenMP.
Kokkos
allows programmers to write modern C++ applications in a hardware-agnostic manner~\cite{DBLP:journals/jpdc/EdwardsTS14}.
It is a programming model for parallel algorithms that use many-core chips and share memory among those cores.
This programming model includes computation abstractions for frequently used parallel computing patterns, policies that provide details for how those computing patterns are applied, and execution spaces that denote on which cores the parallel computation is performed.



Other similar STL-based programming models include
Microsoft's PPL (Parallel Patterns Library)~\cite{ppl_url},
RAJA~\cite{DBLP:conf/ppopp/BeckingsaleHSV19}, etc.
This kind of programming model implements the functions and their extensions in the standard C++ template library,
and provides concurrent data structures.
Thus, this programming model can unbind the parallel programming itself with the underlying hardware resources.
Programmers do not need to care about the details of the underlying architectures,
which effectively lowers the programming barrier.

\subsubsection{Directive-based Programming Models}
Another high-level programming models are based on directive annotations,
including both industry standardards (OpenMP~\cite{omp_url}, OpenACC~\cite{oacc_url}, Intel LEO~\cite{mpss_url}) and
academia-maintained efforts (OmpSs~\cite{DBLP:journals/ppl/DuranABLMMP11},
XcalableMP~\cite{DBLP:conf/ipps/NomizuTLBS12}, Mint~\cite{DBLP:conf/ics/UnatCB11}, OpenMDSP~\cite{DBLP:conf/IEEEpact/HeCCZTY11}).
These programming models only have to add directive constructs before the target code region,
while the tasks of offloading, parallelization and optimization are delegated to compilers and runtime systems.
On one hand, programmers do not have to master a large number of architectural details, thus leading to an improved productivity.
On the other hand, programmers can annotate their codes incrementally, which also lowers the barrier of debugging.
Therefore, this programming model enables the non-experts to enjoy the performance benefits of
heterogeneous many-cores without being entangled in the architecture details.

Multiple directive-based programming models can be mapped onto a single many-core architecture.
As we have mentioned in Section~\ref{subsec:vendor-ppm}, programming the Cell/B.E. processor is challenging.
There is a significant amount of research in programming models that
attempts to make it easy to exploit the computation power of the CEBA architecture.
Bellens \etal present Cell superscalar (CellSs)
which addresses the automatic exploitation of the functional parallelism of a sequential program
through the different processing elements of the CEBA architecture~\cite{DBLP:conf/sc/BellensPBL06}.
Based on annotating the source code, a source-to-source compiler generates the necessary code and a runtime library
exploits the existing parallelism by building a task dependency graph.
The runtime takes care of task scheduling and data movements between the different processors of this architecture.
O'Brien \etal explore supporting OpenMP on the Cell processor~\cite{DBLP:journals/ijpp/OBrienOSCZ08} based on IBM's XL compiler,
so that programmers can continue using their familiar programming model and existing code can be re-used.

\subsubsection{Domain-Specific Programming Models}
To achieve an even better performance and programmability, the domain-specific programming models are preferred on heterogeneous many-cores.

Mudalige \etal propose a high-level programming framework, \texttt{OPS},  
for multi-block structural grid applications~\cite{ops_tr}. 
The frontend of \texttt{OPS} leverages a common set of APIs for users, 
and the backend generates highly optimized device code for target platforms. 
The \texttt{OP2} framework is built upon \texttt{OPS}.
The difference is that \texttt{OPS} is suitable for dealing with multi-block structured grids, 
while \texttt{OP2} is targeted for unstructured grid applications~\cite{DBLP:journals/sigmetrics/GilesMSMK11}.
\texttt{AMGCL} is a header-only C++ library for solving large sparse linear systems with algebraic multigrid (AMG) method~\cite{Demidov2019}. 
This library has a minimal dependency, and provides both shared-memory and distributed memory versions of the algorithms. 
It allows for transparent acceleration of the solution phase with OpenCL, CUDA, or OpenMP~\cite{denis_demidov_2018_1244532}. 

Dubach \etal propose a new Java compatible object-oriented programming language (\texttt{Lime})~\cite{DBLP:conf/pldi/DubachCRBF12}. 
It uses high-level abstractions to explicitly represent parallelism and computation. 
The backend compiler and runtime can automatically manage the data mapping and generate OpenCL/CUDA code for GPUs. 
\texttt{Halide} is a new language for generating efficient image processing code 
on heterogeneous platforms and simplifying programming~\cite{DBLP:conf/pldi/Ragan-KelleyBAPDA13, DBLP:journals/tog/Ragan-KelleyAPLAD12}. 
Its frontend is embedded in C++, while its backend includes x86, ARMv7, CUDA and OpenCL. 
Equally used in image processing, 
\texttt{KernelGenius} is a high-level programming tool 
for EMM (explicitly managed memory many cores)~\cite{DBLP:conf/cases/LepleyPF13}. 
\texttt{Membarth} has implemented a source-to-source compiler, which translates a high-level description 
into low-level GPU codes (OpenCL, CUDA or renderscript)~\cite{DBLP:conf/ipps/MembarthHTKE12, DBLP:journals/tpds/MembarthRHTKE16}. 
Sidelnik \etal have proposed to implement a high-level programming language, \texttt{Chapel}, 
for controlling task allocation, communication and data locality structure on multi-core CPUs and GPUs.
A program in \texttt{Chapel} can run on multiple platforms, and achieve the same performance as CUDA programs~\cite{DBLP:conf/ipps/SidelnikMCGP12}. 
Hong \etal describe \texttt{Green-Marl}, a domain-specific language,
whose high-level language constructs allow developers to describe their graph analysis algorithms intuitively, 
but expose the data-level parallelism inherent in the algorithms~\cite{DBLP:conf/asplos/HongCSO12}. 

The deep learning frameworks (e.g., \texttt{Tensorflow}~\cite{DBLP:journals/corr/AbadiABBCCCDDDG16}, \texttt{PyTorch}~\cite{DBLP:conf/nips/PaszkeGMLBCKLGA19}, 
and \texttt{MXNet}~\cite{DBLP:journals/corr/ChenLLLWWXXZZ15}) 
provide users with script or functional languages (e.g., \textit{Python}, \textit{R}, \textit{Scala}, \textit{Julia}) in the frontend.
These script languages are used to describe the workflow of training or inference.
At the backend, the frameworks dispatch tasks to the underlying heterogeneous systems (GPUs or FPGAs) 
via low-level or other high-level programming models such as OpenCL, CUDA or SYCL.
This whole process of mapping tasks is transparent to users.

To sum up, the domain-specific programming models have the potential to improve programmer productivity, 
to support domain-specific forms of modularity, and to use domain-specific information to support optimizations~\cite{DBLP:journals/tog/MarkGAK03}.
Most of these advantages are obtained by raising the language's abstraction level with domain-speciﬁc data types, operators, and control constructs.
Although the domain-specific programming models can generate efficient kernel code, they are limited to specific application domains. 

\section{Compiling Techniques for Improved Programmability and Portability} \label{sec:bridge}
Translating code in one programming model to another enables code reuse and reduce the learning curve of a new computing language.
Ultimately, code translation can improve programmability, portability, and performance.
In this section, we review the code translation techniques between parallel programming models on heterogeneous many-core architectures.


\vspace{-2mm}
\subsection{C-to-CUDA}
CUDA provides a multi-threaded parallel programming model,
facilitating high performance implementations of general-purpose computations on GPUs.
However, manual development of high-performance CUDA code still requires a large amount of effort from programmers.
Programmers have to explicitly manage the memory hierarchy and multi-level parallel view.
Hence the automatic transformation of sequential input programs into parallel CUDA programs is of significant interest.


Baskaran \etal describe an automatic code transformation system that generates parallel CUDA code from input sequential C code,
for affine programs~\cite{DBLP:conf/cc/BaskaranRS10}.
Using publicly available tools that have made polyhedral compiler optimization practically effective,
the authors develop a C-to-CUDA transformation system that generates two-level parallel CUDA code
that is optimized for efficient data access.
The performance of the automatically generated CUDA code is close to hand-optimized versions
and considerably better than their performance on multi-core CPUs.
Building on Baskaran's experience, Reservoir Labs developed its own compiler based on R-Stream~\cite{DBLP:conf/asplos/LeungVMBWBL10},
which introduces a more advanced algorithm to exploit the memory hierarchy.

\cparagraph{PPCG} Verdoolaege \etal address the compilation of a sequential program for parallel execution
on a modern GPU~\cite{DBLP:journals/taco/VerdoolaegeJCGTC13}.
They present a source-to-source compiler (\texttt{PPCG}),
which singles out for its ability to accelerate computations from any static control loop nest,
generating multiple CUDA kernels when necessary.
The authors introduce a multilevel tiling strategy and
a code generation scheme for the parallelization and
locality optimization of imperfectly nested loops,
managing memory and exposing concurrency according to the constraints of modern GPUs.

\cparagraph{Bones}
Nugteren \etal evaluate a number of C-to-CUDA transformation tools targeting GPUs,
and identify their strengths and weaknesses~\cite{DBLP:conf/asplos/NugterenC12}.
Then they address the weaknesses by presenting a new classification of algorithms.
This classification is used in a source-to-source compiler (\texttt{Bones}) based on the algorithmic skeletons technique.
The compiler generates target code based on skeletons of parallel structures,
which can be seen as parameterisable library implementations for a set of algorithm classes.
This compiler still requires some modifications to the original sequential source code, but
can generate readable code for further fine-tuning.

\cparagraph{PIPS} Non-polyhedral tools have also seen major developments. PIPS
is an open-source initiative developed by the HPC Project to unify
efforts concerning compilers for parallel architectures~\cite{pips_url, DBLP:conf/lcpc/AminiCIK11}.
It supports the automatic integrated compilation
of applications for heterogeneous architectures including GPUs.
The compiler
uses abstract interpretation for array regions based on polyhedra,
which allows PIPS to perform powerful interprocedural analysis on the input code.

\cparagraph{DawnCC}
Mendonca \etal argue that inserting pragmas into production code is a difficult and error-prone task,
often requiring familiarity with the target program~\cite{DBLP:journals/taco/MendoncaGAPAP17, DBLP:conf/cgo/MishraKC19}.
This difficulty restricts developers from annotating code
that they have not written themselves.
Therefore, they provide a suite of compiler-based methods and a tool, \texttt{DawnCC}, to mitigate the issue.
The tool relies on symbolic range analysis
to achieve two purposes: populate source code with data transfer primitives and
to disambiguate pointers that could hinder automatic parallelization due to aliasing.

\vspace{-2mm}
\subsection{CUDA-to-OpenCL}
Restricted to NVIDIA GPUs, CUDA has the largest code base and high-performance libraries.
On the other hand, OpenCL is an open standard supported on a large number of mainstream devices.
With the great interest in OpenCL comes a challenge:
manufacturers have a large investment in CUDA codes and yet would
like to take advantage of wider deployment opportunities afforded by OpenCL.
Therefore, an automated tool for CUDA-to-OpenCL translation is required.

\cparagraph{SnuCL-Tr}
Kim \etal present similarities and differences between CUDA and OpenCL, and develop an automatic translation framework
between them~\cite{DBLP:conf/sc/KimDJJL15}.
\texttt{SnuCL-Tr}
can achieve comparable performance between the original and target applications in both directions.
Given that each programming model has a large user-base and code-base,
this translator is useful to extend the code-base for each programming model and
unifies the efforts to develop applications.

\cparagraph{CU2CL}
Gardner \etal summarize the challenges of translating CUDA code to its OpenCL equivalence~\cite{DBLP:journals/pc/GardnerSFM13}.
They develop an automated CUDA-to-OpenCL source-to-source translator (\texttt{CU2CL}),
to automatically translate three medium-to-large,
CUDA-optimized codes to OpenCL, thus enabling the codes to run on other GPU-accelerated systems~\cite{DBLP:conf/hpcasia/SathreGF19, DBLP:conf/icpads/MartinezGF11}.
\texttt{Swan} is tool used to ease the transition between OpenCL and CUDA~\cite{DBLP:journals/cphysics/HarveyF11}.
Different from CU2CL, Swan provides a higher-level library that abstracts both CUDA and OpenCL,
such that an application makes calls to Swan and Swan
takes care of mapping the calls to CUDA or OpenCL.

\cparagraph{NMT}
Kim \etal present source-to-source translation between CUDA to OpenCL using neural machine translation (\texttt{NMT}).
To generate a training dataset, they extract CUDA API usages from CUDA examples and write corresponding OpenCL API usages.
With a pair of API usages acquired, they construct API usage trees that help users find unseen usages
from new samples and easily add them to a training input~\cite{DBLP:conf/cgo/KimK19}.

\cparagraph{O2render}
With a similar goal, Yang \etal introduces \texttt{O2render}, an OpenCL-to-Renderscript
translator that enables the porting of an OpenCL application to
a Renderscript application~\cite{DBLP:conf/estimedia/YangWL12}.
O2render automatically translates
OpenCL kernels to a Renderscript kernel.

\vspace{-2mm}
\subsection{Directive-to-CUDA/OpenCL}


Translating OpenMP-like codes into CUDA/OpenCL codes will not only reuse the large OpenMP code base,
but also lower their programming barrier.

\cparagraph{OpenMP-to-CUDA}
Lee and Eigenmann present a framework for automatic source-to-source
translation of standard OpenMP applications into CUDA applications~\cite{DBLP:conf/ppopp/LeeME09}.
This translator aims to further improve programmability and make existing OpenMP applications
amenable to execution on GPGPUs.
Later, they propose a new programming interface, \texttt{OpenMPC},
which builds on OpenMP to provide an abstraction of CUDA and
offers high-level controls of the involved parameters and optimizations~\cite{DBLP:conf/sc/LeeE10}.

\cparagraph{OpenMP-to-OpenCL}
Kim \etal propose a framework that translates OpenMP 4.0 accelerator directives to OpenCL~\cite{DBLP:conf/sc/KimLPL16}.
They leverage a run-time optimization to automatically eliminates unnecessary data transfers
between the host and the accelerator.

\cparagraph{OpenMP-to-LEO}
Managing data transfers between the CPU and XeonPhi and optimizing applications for performance
requires some amount of effort and experimentation.
Ravi \etal present \texttt{Apricot}, an optimizing compiler and productivity tool for Intel XeonPhi
that minimizes developer effort by automatically inserting LEO clauses~\cite{DBLP:conf/ics/RaviYBC12}.
This optimizing compiler aims to assist programmers in porting existing multi-core applications and
writing new ones to take full advantage of the many-core accelerator,
while maximizing overall performance.

\cparagraph{CUDA-lite}
CUDA programmers shoulder the responsibility of managing the code to produce the desirable access patterns.
Experiences show that such responsibility presents a major burden on the programmer, and
this task can be better performed by automated tools.
Ueng \etal present \texttt{CUDA-lite}, an enhancement to CUDA, as one such tool~\cite{DBLP:conf/lcpc/UengLBH08}.
This tool leverages programmers' knowledge via annotations to perform transformations and
show preliminary results that indicate auto-generated code can have performance comparable to hand-crafted codes.

\cparagraph{hiCUDA}
Han \etal have designed \texttt{hiCUDA}~\cite{DBLP:journals/tpds/HanA11, DBLP:conf/asplos/HanA09},
a high-level directive-based language for CUDA programming.
They develop a prototype compiler to facilitate the translation of a hiCUDA program to a CUDA program.
The compiler is able to support real-world applications that span multiple procedures and use dynamically allocated arrays.


\cparagraph{CUDA-CHiLL} The CHiLL developers extended their compiler to generate GPU code
with \texttt{CUDA-CHiLL}~\cite{DBLP:conf/lcpc/RudyKHCC10},
which does not perform an automatic
parallelization and mapping to CUDA but instead offers high-level constructs that
allow a user or search engine to perform such a transformation.

\subsection{Adapting CUDA/OpenCL to Multi-core CPUs}
\cparagraph{MCUDA} is a source-to-source translator
that translates CUDA to multi-threaded CPU code~\cite{DBLP:conf/lcpc/StrattonSH08}.
This translator is built on Cetus~\cite{DBLP:journals/ijpp/BaeMLALDEM13},
a source-to-source translator framework for C and other C-based languages.

\cparagraph{CUDA-x86} by PGI allows developers using CUDA to compile and optimize their CUDA applications
to run on x86-based multi-core architectures~\cite{cudax64_url}.
\texttt{CUDA-x86} includes full support for NVIDIA's CUDA C/C++ language for GPUs.
When running on x86-based systems without a GPU,
CUDA applications can use multiple cores and the streaming SIMD capabilities of Intel and AMD CPUs for parallel execution.

\cparagraph{Ocelot}~\cite{DBLP:conf/IEEEpact/DiamosKYC10} is primarily a PTX-to-LLVM translator and
run-time system that can decide whether to run the PTX on a GPU device or on a CPU with just-in-time (JIT) compilation.
\texttt{Ocelot} is similar to \texttt{MCUDA} in that it allows for CUDA kernels to be run on CPUs,
but it takes the approach of performing translations on lower-level bytecodes.

\vspace{-2mm}
\subsection{Supporting Multiple Devices}


The interest in using multiple 
accelerating devices
to speed up applications 
has increased in recent years. However, the existing heterogeneous programming models (e.g., OpenCL) 
abstract details of devices at the per-device level and 
require programmers to explicitly schedule their kernel tasks on a system 
equipped with multiple devices. 
This subsection examines the software techniques of extending parallel programming models to support multiple devices. 


\cparagraph{GPUSs} 
The GPU Superscalar (\texttt{GPUSs}) is an extension of the Star Superscalar (\texttt{StarSs}) programming model 
that targets application parallelization on platforms 
with multiple GPUs~\cite{DBLP:conf/europar/AyguadeBILMQ09}. 
This framework deals with architecture heterogeneity and 
separate memory spaces, 
while preserving simplicity and portability. 

\cparagraph{VirtCL}
You \etal propose a framework (\texttt{VirtCL}) that reduces the programming burden by acting as a layer 
between the programmer and the native OpenCL run-time system.
\texttt{VirtCL} abstracts multiple devices into a single virtual device~\cite{DBLP:conf/ppopp/YouWTC15}.
This framework comprises two main software components: a front-end library, 
which exposes primary OpenCL APIs 
and the virtual device, and  a back-end run-time system (CLDaemon) for 
scheduling and dispatching kernel tasks based on a history-based scheduler. 

\cparagraph{OpenMP extension}
Yan \etal explore support of multiple accelerators in high-level programming models 
by extending OpenMP to support offloading data and computation regions 
to multiple accelerators~\cite{DBLP:conf/ppopp/0001LLSQ15}. 
These extensions allow for distributing data and computation among a list of devices via easy-to-use interfaces, 
including specifying the distribution of multi-dimensional arrays and 
declaring shared data regions. 

\cparagraph{OpenACC-to-CUDA}                                                                                                                                                     
Komoda \etal present an OpenACC compiler 
to run single OpenACC programs on multiple GPUs~\cite{DBLP:conf/icpp/KomodaMNM13}.
By orchestrating the compiler and the runtime system, 
the proposed system can efficiently manage the necessary data movements 
among multiple GPUs memories. 
The authors extend a set of directives based on the standard OpenACC API to facilitate communication optimizations. 
The directives allow programmers to express the patterns of memory accesses in the parallel loops to be offloaded. 
Inserting a few directives into an OpenACC program can reduce a large amount of unnecessary data movements.

\section{Optimization Techniques for Minimizing the Host-Accelerator Communication} \label{sec:stream}
While the heterogeneous many-core design offers the
potential for energy-efficient, high-performance computing,
software developers are finding it increasingly hard to deal
with the complexity of these systems~\cite{DBLP:journals/ijhpca/MeswaniCUSBP13, DBLP:conf/icpp/FangVS11}.
In particular, programmers need to effectively manage the host-device
communication, because the communication overhead can
completely eclipse the benefit of computation offloading
if not careful~\cite{DBLP:conf/ispass/GreggH11, DBLP:conf/iiswc/CheBMTSLS09, DBLP:conf/ipps/BoyerMK13, DBLP:conf/wosp/FangSZXCV14}.
Gregg and Hazelwood have
shown that, when memory transfer times are included,
it can take 2x--50x longer to run a kernel than the GPU processing time alone~\cite{DBLP:conf/ispass/GreggH11}.

Various parallel programming models have introduced the streaming mechanism to
amortize the host-device communication cost~\cite{DBLP:conf/ipps/NewburnBWCPDSBL16}.
It works by partitioning the processor
cores to allow independent communication and computation
tasks (i.e., streams) to run concurrently on different hardware
resources, which effectively overlaps the kernel execution
with data movements.
Representative heterogeneous streaming implementations include CUDA Streams~\cite{cuda_url}, OpenCL
Command Queues~\cite{ocl_url}, and Intel's hStreams~\cite{tr:hstreams:arch, DBLP:conf/ipps/NewburnBWCPDSBL16, DBLP:conf/npc/LiFTCY16, DBLP:journals/ppl/FangZLTCCY16, DBLP:conf/ipps/ZhangFTYW18, DBLP:conf/ipps/LiFTCCY16}.
These implementations allow the program to spawn more
than one stream/pipeline so that the data movement stage of
one pipeline overlaps the kernel execution stage of another.

\subsection{The Streaming Mechanism}
The idea of heterogeneous streams is to exploit temporal and spatial sharing of the computing resources.

\begin{figure}[!t]
\centering
\includegraphics[width=0.45\textwidth]{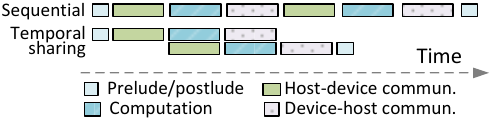}
\caption{Exploit pipeline parallelism by temporal sharing. Reproduced from \cite{DBLP:conf/ipps/ZhangFTYW18}.}
\label{fig:temporal_sharing}
\end{figure}

\cparagraph{Temporal Sharing.} Code written for heterogeneous computing devices typically consists of several stages such as host-device
communication and computation. Using temporal sharing, one can overlap some of these stages to exploit pipeline parallelism to improve
performance. This paradigm is illustrated in Figure~\ref{fig:temporal_sharing}. In this example, we can exploit temporal sharing to overlap
the host-device communication and computation stages to achieve better runtime when compared to  execute every stage sequentially.
One way of exploiting temporal sharing is to divide an application into independent tasks so that they can run in a pipeline fashion.

Note that the PPE and SPEs of Cell/B.E. share the same memory space, and
thus there is no need of host-accelerator communication~\cite{DBLP:journals/ibmrd/ChenRDI07}.
But each SPE has a channel/DMA transport for controlling input and output.
To prevent memory stalls, we can take advantage of the DMA engines by
adopting a double-buffer approach to overlap computations on the previously fetched datablocks
with transfers of subsequent data blocks to and from memory.
We regard that this is a special case of \textit{temporal sharing}.

\begin{figure}[!t]
\centering
\includegraphics[width=0.45\textwidth]{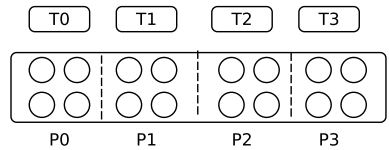}
\caption{Spatial sharing. The circles represent processing cores, \texttt{Tx} represents a task, and \texttt{Px} represents a partition. }
\label{fig:spatial_sharing}
\end{figure}

\cparagraph{Spatial Sharing.}
Using multiple streams also enjoys the idea of resource partitioning. That is, to partition the resource (e.g., processing cores) into multiple groups and map each stream onto a partition. Therefore, different streams can run on different partitions concurrently, i.e., resource \textit{spatial sharing}. Nowadays accelerators have a large number of processing units that some applications cannot efficiently exploit them for a given task. Typically, we offload a task and let it occupy all the processing cores. When using multiple streams, we divide the processing cores into multiple groups (each group is named as a \textit{partition}). Figure~\ref{fig:spatial_sharing} shows that a device has 16 cores and is logically divided into four partitions (\texttt{P0}, \texttt{P1}, \texttt{P2}, \texttt{P3}). Then different tasks are offloaded onto different partitions, e.g., \texttt{T0}, \texttt{T1}, \texttt{T2}, \texttt{T3} runs on \texttt{P0}, \texttt{P1}, \texttt{P2}, \texttt{P3}, respectively. In this way, we aim to improve the device utilization.

\vspace{-4mm}
\subsection{Intel hStreams}
Intel \texttt{hStreams} is an open-source streaming implementation built on the COI programming model~\cite{DBLP:conf/ipps/NewburnBWCPDSBL16}.
At its core is the resource partitioning mechanism~\cite{tr:hstreams:arch}.
At the physical level, the whole device is partitioned into multiple groups and thus each group has several processing cores. At the logical level, a device can be seen as one or more \textit{domains}. Each domain contains multiple \textit{places}, each of which then has multiples \textit{streams}. The logical concepts are visible to programmers, while the physical ones are transparent to them and the mapping between them are automatically handled by the runtime.

\texttt{hStreams} is implemented as a library and provides users with APIs to access coprocessors/accelerators efficiently.
Programming with \texttt{hStreams} resembles that in \texttt{CUDA} or \texttt{OpenCL}.
Programmers have to create the streaming context, move data between host and device, and invoke kernel execution.
And they also have to split tasks to use multiple streams.
Figure~\ref{fig:example_code} gives a simplified code example written with Intel's \texttt{hStreams} APIs.
At line 2, we initialize the stream execution by setting the number of partitions and tasks/streams per partition.
This initialization process essentially creates multiple processor domains and
determines how many logical streams can run on a partition.
In the \emph{for} loop (lines 7-14) we enqueue the communication and computation tasks to a number of streams identified by the
\texttt{stream\_id} variable. In this way, communication and computation of different streams can be overlapped during execution
(temporal sharing); and streams on different processor domains (or partitions) can run concurrently (spatial sharing).

\lstset{}
\begin{figure}[!t]
	\centering %
		\noindent\mbox{\parbox{\columnwidth}{%
\begin{lstlisting}[label=subfig:source_in]
//setting the partition-size and task granularity
hStreams_app_init(partition_size,streams_p_part);

//stream queue id
stream_id = 0;
for(...){
  //enquque host-device transfer to current stream
  hStreams_app_xfer_memory(,,, stream_id, HSTR_SRC_TO_SINK,...);
  ...
  //enqueue computation to the current stream
  hStreams_EnqueueCompute(stream_id, "kernel1", ...);
  ...
  //move to the next stream
  stream_id = (stream_id++) % MAX_STR;
}
//transfer data back to host
hStreams_app_xfer_memory(,,, HSTR_SINK_TO_SRC,...);
\end{lstlisting}%
		}}
    \vspace{-3mm}
	\caption{Example \texttt{hStreams} code.}%
	\label{fig:example_code}%
\end{figure}
\subsection{Performance Modelling for Streaming Programs}
The previous work have demonstrated that choosing a right stream configuration
has a great impact on the resultant performance~\cite{DBLP:conf/ipps/ZhangFTYW18, DBLP:journals/ppl/FangZLTCCY16}.
And the best configuration must be determined on a per-program and per-dataset basis.
Attempting to find the optimal configuration through
an exhaustive search would be ineffective,
and the overhead involved would be far bigger than the potential
benefits. Therefore, building models for heterogeneous streaming programs is of great significance.

\subsubsection{Hand-Crafted Analytical Models}
\vspace{-2mm}
Gomez-Luna \etal present performance models for asynchronous data transfers on
 GPU architectures~\cite{DBLP:journals/jpdc/Gomez-LunaGBG12}.
The models permit programmers to estimate the optimal number of streams in which the computation on the GPU should be broken up.
Werkhoven \etal present an analytical performance model to indicate when to
apply which overlapping method on GPUs~\cite{DBLP:conf/ccgrid/WerkhovenMSB14}.
The evaluation results show that the performance model is capable of correctly classifying the relative performance of the different implementations.
Liu \etal carry out a systematic investigation into task partitioning to
achieve maximum performance gain for AMD and NVIDIA GPUs~\cite{DBLP:journals/ijhpca/LiuQJG16}.
This approach is not ideal, as it is not only complex to develop the analytical models, but is likely to fail
due to the variety of programs and the ever-changing hardware architecture.
That is, these hand-crafted models have the drawback of being not
portable across architectures as the model is tightly coupled to a specific many-core architecture.


\subsubsection{Machine-Learning based Models}
\vspace{-2mm}
Researchers have also exploited the machine learning techniques to automatically construct a predictive model to
directly predict the best configuration~\cite{DBLP:conf/ipps/ZhangFTYW18, tpds20}.
This approach provides minimal runtime, and
has little development overhead when targeting a new many-core architecture.
This is achieved by employing machine learning techniques to automatically construct a predictive model to decide at
runtime the optimal stream configuration for
any streamed application.
The predictor is first trained \emph{off-line}. Then, using code and dynamic runtime features of the program,
the model predicts the best configuration for a \emph{new}, \emph{unseen} program.
This approach can avoid the pitfalls of using a hard-wired
heuristic that requires human modification every time when the architecture evolves,
where the number and the type of cores are likely to change from one generation to the next.
Experimental results XeonPhi and GPGPUs have shown that this approach can achieve over 93\% of the Oracle performance~\cite{tpds20}.

\section{A Vision for the Next Decade} \label{sec:next}
Given the massive performance potential of heterogeneous many-core hardware design, it is clear that future computing hardware will be
increasingly specialized and diverse. As the hardware design keeps evolving, so does the software development systems and the programming
model. In this section, we discuss some challenges and open research directions for future parallel programming models.

\cparagraph{A holistic solution.} China, US, Japan and EUROPE are currently working towards the exascale computing.
The design and construction of an exascale machine will be built based on
heterogeneous many-core architectures of various forms~\cite{DBLP:journals/jzusc/LiaoLYLYLHLFRS18}.
This achievement will require significant advances in the software paradigm
and require that parallelism in control and data be exploited at all possible levels.
Therefore,
the dominant design parameter
will shift from hardware to system software and
in particular, parallel programming systems~\cite{DBLP:journals/tjs/DavisRFNT12}.
We envision that a hierarchy of programming models have to be implemented as well as the equipment of expert optimized libraries.
Low-level programming models should be implemented, but are not suggested to be exposed to programmers.
Instead, the high-level programming models and highly optimized domain libraries are exposed as the programming interface.
Intel's OneAPI is one of such examples~\cite{oneapi_url}.

\cparagraph{Pattern-aware parallel programming.} Parallel programming based on patterns (such as \textit{map/reduce} and \textit{pipeline})
or algorithmic skeletons (Section \ref{sec:skeleton}), where programmers write algorithmic intents that abstract away parallelization,
heterogeneity, and reliability concerns, offer a partial solution for programming heterogeneous parallel systems~\cite{de2017bringing}. As
can been from Section~\ref{sec:highlevel-ppm}, this is an essential feature of high-level parallel programming models. The key to the
success of pattern-based parallel programming is to have a fully-supported development toolchain for code optimization, debugging and
performance diagnosis. The current compiler implementation is oblivious to the high-level algorithmic intents expressed in the patterns,
leading to disappointing performance, discouraging the adoption of pattern-based parallel programming. For example, a sequential loop that
adds one to an integer literal one million times will be optimized away at compile time. However, if we implement it as a parallel pipeline
pattern, existing compilers, including leading industrial parallel compilers, Intel TBB on ICC and Go, would fail to merge the trivial
pipeline elements. As a result, the pattern-based implementation takes minutes to run, not nanoseconds\footnote{Code is available at:
\url{https://goo.gl/y7bBdN}.}, leading to a massive slowdown over the sequential version. The issue arises from the fact that current
compilers are oblivious to parallel patterns. A parallel construct encodes the sequential semantics of the program, but this is lost to the
compiler. If the compiler knew the sequential semantics, it can then dynamically merge small pipeline elements. If we can do these, the
primary barrier to adopting pattern-based programming would be torn down.


\cparagraph{Machine learning based program tuning.} Programmers are faced with many decisions when writing heterogeneous programs, such as
selecting an optimal thread configuration~\cite{DBLP:journals/tpds/DaoL18} and/or selecting a right code
variant~\cite{DBLP:conf/asplos/MuralidharanRHG16}. This is due to the profound differences in many-core architectures, programming models
and input applications~\cite{DBLP:journals/pieee/BalaprakashDGHH18}. By default, the runtime system of high-level programming models has to
assist users in automating these online decisions. If a wrong configuration is selected, the overall performance will be significantly
decreased. Therefore, it is significant to design a model to help programmers to automatically choose a reasonable configuration, i.e.,
\textit{automated performance tuning}, which is regarded to have the potential to dramatically improve the performance portability of
petascale and exascale applications~\cite{DBLP:journals/pieee/BalaprakashDGHH18}. A key enabling technology for optimizing parallel
programs is \emph{machine learning}. Rather than hand-craft a set of optimization heuristics based on compiler expert insight, learning
techniques automatically determine how to apply optimizations based on statistical modelling and learning~\cite{zhengphd,mlcpieee}. This
provides a rigorous methodology to search and extract structure that can be transferred and reused in unseen settings.  Its great advantage
is that it can adapt to changing platforms as it has no a priori assumption about their behaviour. There are many studies showing it
outperforms human-based approaches. Recent work shows that it is effective in performing parallel code
optimization~\cite{grewe2013portable,DBLP:journals/taco/WangGO14,cc14,ogilvie2014fast,cgo17,cummins2017end,chen2020characterizing},
performance predicting~\cite{zhao2016predicting,wang2013using}, parallelism
mapping~\cite{Tournavitis:2009:THA:1542476.1542496,wang2010partitioning,grewe2013opencl,DBLP:journals/taco/WangGO14,wen2014smart,wang2014integrating,taylor2017adaptive,tpds20},
and task scheduling~\cite{emani2013smart,ren2017optimise,conext18,yuan2019using,ren2020camel,sanz2019optimizing}. As the many-core design
becomes increasingly diverse, we believe that the machine-learning techniques provide a rigorous, automatic way for constructing
optimization heuristics, which is more scalable and sustainable, compared to manually-crafted solutions.

\section{Conclusions}
 This article has introduced programming models for heterogeneous many-core architectures. Power, energy and thermal limits have
forced the hardware industry to introduce heterogeneous many-cores built around specialized processors. However, developers are struggling
to manage the complexity of heterogeneous many-core hardware. A crisis is looming as these systems become pervasive. As such, how to enable
developers to write and optimize programs for the emerging heterogeneous many-core architectures has generated a large amount of academic
interest and papers. While it is impossible to provide a definitive cataloger of all research, we have tried to provide a comprehensive and
accessible survey of the main research areas and future directions. As we have discussed in the article, this is a trustworthy and exciting
direction for systems researchers.

\begin{acknowledgements}
This work was partially funded by the National Key Research and Development Program of China under Grant No. 2018YFB0204301, the National
Natural Science Foundation of China under Grant agreements 61972408, 61602501 and 61872294, and a UK Royal Society International
Collaboration Grant.
\end{acknowledgements}

%
%

\balance
\bibliographystyle{spmpsci}      
\bibliography{ref,zheng}   


\end{document}